\documentstyle[aps,psfig,epsf]{revtex}

\input epsf

\tightenlines

\begin{document}

{\Large \bf Corrigendum}

\vspace{0.4in}

{\bf The Quark-Gluon-Plasma Liquid}

Markus H. Thoma

2005 {\it J. Phys. G: Nucl. Phys.} {\bf 31} L7-L12 


\vspace{0.4in}

The estimate of the Coulomb coupling parameter $\Gamma $ contains an error. 
In QCD, where the Heavyside-Lorentz units are used, the Coulomb potential
has to be divided by a factor $4 \pi$ compared to CGS units \cite{Jackson}.
Hence the correct Coulomb coupling parameter reads $\Gamma = Cg^2/(4\pi dT)$.
Taking into account the magnetic interaction which is of the same magnitude
as the electric interaction in an ultrarelativistic plasma, the coupling
parameter is reduced by about a factor of 6. Consequently we obtain 
$\Gamma = 1.5$-5 in a QGP at $T\simeq 200$ MeV. Such a value still indicates
that the QGP is in the liquid phase. However, the phase transition to
the gas phase, assumed to happen at $\Gamma_c \simeq 1$, takes place now at
a few times of the transition temperature from the hadronic phase to the QGP.
Hence it might be possible that the gas-liquid transition occurs during the
expansion of the fireball in nucleus-nucleus collisions at LHC \cite{Peshier}.
The phase transition from the QGP liquid to the QGP gas expands the phase
diagram of strongly interacting matter to high temperatures. This phase 
transition ends at a critical point, above which a supercritical fluid exists.
  
The estimate of the cross section enhancement is also affected by the above 
error. The Coulomb radius should also be divided by a factor of 6, leading to 
$\rho = 0.2$-1 fm. Then $\beta = \rho/\lambda_D = 1$-5 which gives a maximum 
impact parameter of (1.4-3.3)$\lambda_D$, from which, using equation (2), a 
cross section enhancement of a factor of 2 to 9 results.    

\bigskip

{\bf Acknowledgments}

\medskip

The author would like to thank C. Greiner and A. Peshier for pointing out
the error in the original article.

\nopagebreak


\end{document}